\documentclass[11pt]{amsart}

\title {Tree Automata and Essential Subtrees}
\author{ Slavcho Shtrakov}
\address{Dept. of Computer  Sciences, South-West University,
 Blagoevgrad }
\email{shtrakov@aix.swu.bg}
\date{}
\usepackage{graphicx}
\usepackage{graphics}
\usepackage{latexsym}
\usepackage{latexsym}
\newtheorem{l00}{\bf Lemma}

\newtheorem{t00}{\bf Theorem}
\newtheorem{e00}{\bf Example}

\newtheorem{d00}{ Definition}
\newtheorem{c00}{\bf Corollary}
\def\Pr{\bf Proof.\ }
\def\fbx{\hfill${}^{\rule{2mm}{2mm}}$}

\def\la{\leftarrow}

\def\ra{\rightarrow}

\def\F{{\mathcal F}}
\def\A{{\mathcal A}}

\relax
\begin{document}

\begin{abstract}
Terms considered as trees are very useful tools for presentation
and manipulation of different data structures in Computer Science
and Information Technology. Any work of the computer over the
trees, for instance as graphical user interface (menus of the
different windows), XML - documents, definitions and description
of hierarchial objects in C++ or Java programming etc. can be
considered as runs of an automaton (the given computer) over these
trees. Often the automaton loses much time to work over a tree
spending it in some useless checking
 of nonessential information. Usually it means that some inputs or
 some subtrees of the given tree are fictive for this automaton.
 The studying  of fictive and essential subtrees of a tree with
 respect to an automaton is main aim of the article.
 We introduce essential subtrees
for terms (trees) and tree automata . There are
some results concerning independent sets of subtrees and
separable sets for a tree and an automaton.
\end{abstract}
\maketitle
 \noindent {\it AMS, subject classification}: 03D05,
68Q70, 03D15, 06B25

\section{Introduction}
Recall that a tree automaton $\A$ is a tuple $\A=( Q, \F, \Delta,
Q_f)$ where: $Q$ is set of states, $Q_f$ - set of final states,
$\Delta$ - set of transition rules and $\F$ is set of operation
symbols.
\\
Tree automata work (run) over trees (terms), composed with operation
symbols from $\F$ and  variables from $X=\{x_1,x_2,\ldots\}.$ An
automaton assigns to the vertices  of tree some states according to
the transition rules from $\Delta.$ A tree is recognized by an
automaton if it assigns to the root of the tree some final state
from $Q_f.$ In section 2 we give the basic definitions and facts
from universal algebra (see \cite{Bu-S;00,Ros,Den3}) and trees (see
\cite{Com,Ges}). In section 3 the concept of essential inputs and
essential subtrees  for a term and an automaton is introduced. An
subtree $t_i$ of $t$ is essential for the tree $t$ and an automaton
$\A$ iff there are two runs of $\A$ over $t$ whose initial states
are different on the inputs belonging to $t_i$ only, and the
resulting states of these runs are different. A generalization of
essential subtrees  is the concept of separable sets of subtrees.
\section{ Trees and Independent Sets}
Let ${\F}$ be any finite set, the elements of which are called $
operation\ symbols. $ Let $\tau:{{\F}}\to N$ be a mapping into the
non negative integers; for $f\in{\F},$ the number $\tau(f)$ will
denote the {\it arity } of the operation symbol $f.$ The pair
$(\F,\tau)$ is called {\it type} or {\it signature}. If it is
obvious what the set ${\F}$ is, we will write "$ type\ \tau$". The
set of symbols of arity $p$ is denoted by ${\F}_p.$ Elements of
arity $0,1,\ldots, p$ respectively are called {\it
constants(nullary), unary,...,$p$-ary} symbols. We assume that
${\F}_0\neq\emptyset.$
\begin{d00}\rm\label{d2}
Let $X_n=\{x_1,\ldots,x_n\}, n\geq 1,$ be a set of variables with
$X_n\cap{{\F}}=\emptyset.$
The set $W_{\tau}(X_n)$ of {\it  $n-$ary terms\ of\ type} $\tau$
with\ variables\ from\ $ X_n $ is defined as the smallest
set for which:
\\
$(i)$\ $\F_0\subseteq W_{\tau}(X_n)$ and
\\
$(ii)$\ $X_n\subseteq W_{\tau}(X_n)$ and
\\
$(iii)$\ if $p\geq 1, f\in\F_p $ and
$t_1,\ldots,t_p\in W_{\tau}(X_n)$ then $f(t_1,\ldots,t_p)\in W_{\tau}(X_n).$
\end{d00}
By $W_\tau(X)$ we denote the following set
$$W_\tau(X):=\cup_{n=1}^{\infty} W_\tau(X_n),$$
where $X=\{x_1,x_2,\ldots \}.$
\\
Let $t$ be a term. By $Var(t)$  the set of all variables from $X$
which occur in $t$ is denoted. The elements of $Var(t)$ are called
{\it input variables (inputs)} for $t$.
 \\
If $t\in W_\tau(X)$ and $s_x\in W_\tau(X)$ then the term denoted
by $t(x\la s_x),$ is obtained by substituting in $t,$
simultaneously for every $x\in X,\quad s_x.$
\\
If $t,s_x\in W_\tau(X_n)$, one may then write $t(x\la s_x)$ in the more
explicit form $t(x_1\la s_{x_1},\ldots,x_n\la s_{x_n}).$
\\
Any subset $L$ of $W_{\tau}(X)$ is called {\it term-language} or
{\it tree-language.}
\\
If $X=\emptyset$ then  $W_\tau(X)$ consists of so called {\it
ground terms}, which can be viewed as usual terms after replacing
in them the all variables with some nullary operation symbols.
\\
Let $t$ be a term of type $\tau.$ The $depth$ of $t$ is defined
inductively: if $t\in X\cup\F_0$ then $Depth(t)=0;$ and if
$t=f(t_1,\ldots,t_{n})$ then
\\
$Depth(t)=max
\{Depth(t_1),\ldots,Depth(t_{n})\} +1.$
\\
Let $N$ be the set of natural numbers and $N^*$ be the set of finite
strings over $N.$  The set $N^*$ is naturally ordered by $\overline
n\preceq \overline m \iff \overline n$\ is a prefix of $\overline
m.$ The empty string in $N^*$ we will denote by $\varepsilon.$
 \\
A term $t\in W_{\tau}(X)$ can also be defined as a partial
function $t:N^*\to {{\F}}\cup X$ with domain $Pos(t)\subset N^*$
satisfying the following properties:
\\
$(i)$\quad $\varepsilon\in Pos(t)$  and $Pos(t)$ is prefix-closed;
\\
$(ii)$\quad For each $\overline p\in Pos(t)$, if $t(\overline p)\in
{\F}_n,$ $n\geq 1$ then $\{i|\overline pi\in Pos(t)\}=
\{1,\ldots,n\};$
\\
$(iii)$\quad For each $\overline p\in Pos(t)$, if $t(\overline p)\in
X\cup\F_0$ then $\{i|\overline pi\in Pos(t)\}=\emptyset.$
\\
The elements of $Pos(t)$ are called  {\it positions.}
\\
A {\it subterm (subtree)} $t|_{\overline p}$ of a term $t\in
W_{\tau}(X)$ at position $\overline p$ is the tree with
$Pos(t|_{\overline p})=\{\overline q\in Pos(t)\ |\ \overline
p\prec\overline q\}.$
\\
The set of all subtrees of $t$ is denoted by $Sub(t).$
\\
 We denote by $\unlhd$ the subterm ordering, i.e. we
write $t\unlhd t'$ if there is a position $\overline p\in Pos(t')$
such that $t=t'|_{\overline p}$ and one says that $t$ is a subterm
of $t'.$ We write $t\lhd t'$ if $t\unlhd t'$ and $t\neq t'.$
\\
Clearly, $t|_{\overline p}\lhd t|_{\overline q}$ if and only if
$\overline q\prec \overline p.$
\\
A chain of subterms $Ch:=t_{p_1}\lhd t_{p_2}\lhd \ldots \lhd
t_{p_k}$ is called {\it strong} if for all $j\in \{1,\ldots,k-1\}$
there does not exist any term $s$ such that $t_{p_j}\lhd s\lhd
t_{p_{j+1}}.$
\begin{l00}\rm\label{le1}
A chain $t|_{\overline{p_1}}\lhd\ldots\lhd t|_{\overline{p_k}}$ is
strong if and only if $\overline{p_{i+1}}\prec \overline{p_i}$ and
there are integers $k_i\in N$ such that
$\overline{p_{i+1}}=\overline{p_i}k_i$ for each $i,
i\in\{1,\ldots,k-1\}.$
\end{l00}
\begin{d00}\rm\label{def11}
Let ${\overline p_1},{\overline p_2}\in Pos(t).$  The positions
${\overline p_1}$ and ${\overline p_2}$ are called {\it independent}
if $Pos(t|_{\overline{p_1}})\cap
Pos(t|_{\overline{p_2}})=\emptyset.$
\end{d00}
If $t_1,t_2\in Sub(t)$ are the subtrees corresponding to the
positions $\overline{p_1},\overline{p_2}\in Pos(t)$ i.e.
$t_1=t|_{\overline{p_1}}$ and $t_2=t|_{\overline{p_2}}$ which are
independent then the subtrees $t_1$ and $t_2$ are called {\it
independent}, too.
\\
The set $Ind(\overline{p_1})$  consists of all positions  $\overline
q \in Pos(t)$ which are independent on $\overline{p_1}$.
Analogously, the set $Ind(t_1)$ consists of all subtrees $s \in
Sub(t)$ which are independent on $t_1$.
\begin{d00}\rm
A set of strings over the set of natural numbers $P$ is called:
\\
$(i)$\quad {\it prefix closed} iff $\overline p\in P$ and $\overline
q\prec \overline p$ imply
 $\overline q\in P;$
\\
$(ii)$\quad {\it prefix determined w.r.t. the set $Q,\ Q\subset
N^*$} iff $\overline p\in P,\ \ \overline p\prec\overline q$ and
$\overline q\in Q$ imply $\overline q\in P.$
\end{d00}
\begin{l00}\rm\label{l2}
For each $\overline p,\ \overline p\in Pos(t)$ the set
$Ind(\overline p)$ is prefix determined w.r.t. $Pos(t).$
\end{l00}
\Pr\rm Let $\overline p\in Pos(t)$ and $\overline q\in Ind(\overline
p).$ Hence $Pos(t|_{\overline p})\cap Pos(t|_{\overline
q})=\emptyset.$ Clearly if $\overline q\prec\overline r$ and
$\overline r\in Pos(t)$ then $\overline r\in Pos(t|_{\overline q}).$
From $Pos(t|_{\overline r})\subset Pos(t|_{\overline q})$ it follows
that $Pos(t|_{\overline r})\cap Pos(t|_{\overline p})=\emptyset$
i.e. $\overline r \in Ind(\overline p).$ \fbx
\begin{e00}\rm\label{examp1}
Let us illustrate all these notion by the following tree:
\\
$t=f_1(g(f_1(x_1,x_2)),f_2(g(f_1(x_3,f_1(x_4,x_3))),g(f_1(x_2,x_1))))$
drown on Figure \ref{f1}.
 \\
1. The set of positions of this tree is $Pos(t):=$
$$:= \{\overline 0,\overline 1,\overline
2,\overline{11},\overline{111},\overline{112},\overline{21},\overline{211},
\overline{2111},\overline{2112},\overline{21121},\overline{21122},
\overline{22},\overline{221},\overline{2211},\overline{2212}\}.$$
 \\
2. The set of independent positions on the position $\overline{111}$
is:
$$Ind(\overline{111})=\{\overline
2,\overline{21},\overline{211},\overline{2111},\overline{2112},\overline{21121},\overline{21122},
\overline{22},\overline{221},\overline{2211},\overline{2212}\}. $$
\\
3. The set of independent subtrees on the tree  $t|_{\overline{2}}$
is:
$$Ind(t|_{\overline{2}})=\{t|_{\overline{1}},t|_{\overline{11}},t|_{\overline{111}},t|_{\overline{112}}\}. $$

\begin{figure}
  \includegraphics[width=12cm]{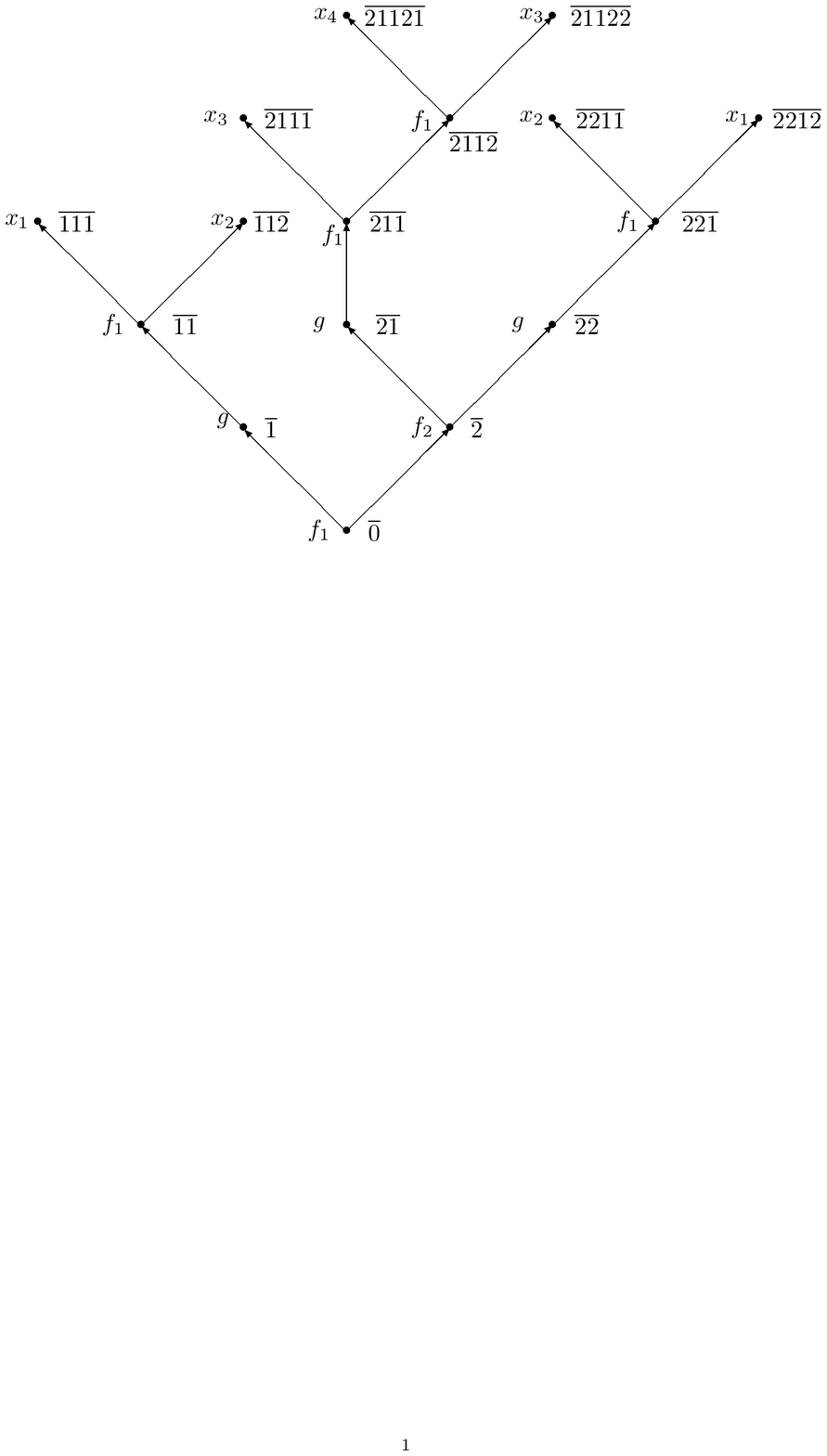}\\
  \caption{~~}\label{f1}
\end{figure}

\end{e00}
\section{   Finite Tree Automata and
Essential Subtrees}
\begin{d00}\rm\label{d5}
A {\it finite tree automaton}  is a tuple
$\A=\langle Q,\F, Q_f,\Delta\rangle$ where:
\\
- $Q$ is a finite set of states;
\\
- $Q_f\subseteq Q$ is a set of final states;
\\
- $\Delta$ is a set of transition rules i.e. if
$$\F= \F_0\cup\F_1\cup\ldots\cup\F_n\quad
{\mbox{\rm then}}\quad \Delta=\{\Delta_0,\Delta_1,\ldots,\Delta_n\},$$
where  $\Delta_i$
are  mappings
$\Delta_0:F_0\ra Q,$
and
$\Delta_i:\F_i\times Q^i\ra Q,$ for $ i=1,\ldots,n.$
\end{d00}
\noindent We will suppose that $\A$ is complete i.e. the
$\Delta$'s are total functions on their domains.
\\
Let $Y\subseteq X$ be a set of input variables. $Ass(Y,\F_0)$
 denotes the set of all functions (evaluations)
$\gamma:Y\ra \F_0.$
\\
Let $t\in W_{\tau}(X),$ $\gamma\in Ass(Y,\F_0)$ and
$Y=\{x_1,\ldots,x_m\}.$ By $\gamma(t)$ the term
$\gamma(t)=t(x_1\la \gamma(x_1),\ldots,x_m\la \gamma(x_m))$
will be denoted.
 \\
So, each assignment $\gamma\in Ass(Y,\F_0)$ can be extended to
a mapping defined on the set
$W_{\tau}(X)$ of all terms.
\\
Let  $ \gamma\in Ass(X,\F_0).$ The automaton $\A=\langle Q,\F,
Q_f,\Delta\rangle$
 runs over $t$ and $\gamma.$ It starts at
leaves of $t$ and moves downwards, associating along a run a
resulting state with each subterm inductively:
\\
If $t=x_i\in X$ then the automaton associates with $t$ the state
$q\in Q,$ with $q=\Delta_0(\gamma(x_i)).$
\\
If $t=f_0\in\F_0$ then the automaton  associates with $t$ the
state   $q=\Delta_0(f_0).$
\\
If $t=f(t_1,\ldots,t_n)$ and the states  $q_1,\ldots,q_n$ are
associated with the subterms(subtrees) $t_1,\ldots,t_n$ then with
$t$ the automaton $\A$ associates the state $q,$ with
$q=\Delta_n(f,q_1,\ldots,q_n).$
\\
The automaton runs  over ground terms  and each assignment from
$Ass(X,F_0)$ transforms any tree as a ground term.
\\
A term $t\ ,t\in W_{\tau}(X)$ is accepted by a tree automaton
$\A=\langle Q,\F, Q_f,\Delta\rangle$ if there exists an assignment
$\gamma$ such that when running over $t$ and $\gamma$ the
automaton $\A$ associates with $t$ a final state  $q\in Q_f.$
\\
When $\A$ associates the state $q$ with a subterm $s,$ we will write
$ \A(\gamma,s)=q.$
\\
If $\gamma\in Ass(Y,\F_0), \ Y\subset X$ then $\A(\gamma,t)$ is
resulting tree obtained from $t$ under applying all possible
transition rules from $\Delta $ over the tree $\gamma(t).$
\\
Let $t\in W_{\tau}(X)$ be a term and $\A$ be a tree automaton which accepts
$t.$ In this case one says that $\A$ {\it recognizes} $t$ or $t$ is
{\it recognizable} by $\A.$ The set of all by $\A$
 recognizable terms is called {\it tree-language} recognized by $\A.$
\begin{d00}\rm\label{d6}
Let $t\in W_{\tau}(X)$ and let $A$ be an automaton. A subtree
$t|_{\overline p},\ \overline p\in Pos(t)$ is called {\it essential}
for the pair $(t,{\A})$ if there exist  two assignments $\gamma_1,
\gamma_2\in Ass(X,{\F}_0)$ such that $\A(\gamma_1,t|_{\overline
p})\neq \A(\gamma_2,t|_{\overline p}),\quad \mbox{\rm and}\quad
\gamma_1(x_j)=\gamma_2(x_j) \quad\mbox{\rm for all}\quad x_j\in
X\setminus Var(t|_{\overline p})$
 with
${\A}(\gamma_1,t)\neq {\A}(\gamma_2,t)$
i.e. ${\A}$ stops in
different states when running over $t$ with  $\gamma_1$ and  with
$\gamma_2.$
\end{d00}
The set of all essential subtrees for the pair $(t,{\A})$ is
denoted by $TEss(t,{\A}).$
 \\
The subtrees from $Sub(t)\setminus TEss(t,{\A})$ are called {\it
fictive } for $(t,{\A}).$
\\
 A position $\overline p\in Pos(t)$ is called
{\it essential } for $(t,\A)$ if $t|_{\overline p}\in TEss(t,{\A}).$
 $PEss(t,{\A})$ is the set
of all essential positions of $t.$
\\
Analogously, the positions from $Pos(t)\setminus PEss(t,{\A})$ are
called {\it fictive } for $(t,{\A}).$
\\
 The set of essential input variables is denoted by
$Ess(t,\A)$. It is studied  in \cite{s1}.
\\
The essential variables for discrete functions are considered in
\cite{Ch,Sa}. Essential input variables and separable sets for terms
in the concept of universal algebra are considered in \cite{sd1}.
\begin{e00}\rm\label{examp2}
Let us consider the automaton defined as follows:
 \\
 $\A=\langle
Q,\F,\Delta,Q_f \rangle,$ where $Q=\{q_0,q_1\};\ \
\F=\F_0\cup\F_1\cup\F_2$ with $\F_0=\{0,1\},\ \ \F_1=\{g\}, \ \
\F_2=\{f_1,f_2\};$ $Q_f=\{q_1\}$ and $\Delta$ is defined by the
equations
 $\Delta_0(0)=q_0,\ \Delta_0(1)=q_1;$\ \
$\Delta_1(g,q_0)=q_1,\ \Delta_1(g,q_1)=q_0;$\ \
$\Delta_2(f_1,q_i,q_j)=q_m,\ \ m=1$ if $q_i=q_j=q_1$ and $m=0$
otherwise;
 $\Delta_2(f_2,q_i,q_j)=q_l,\ \ l=0$ if $q_i= q_j=q_0$ and $l=1$
otherwise.
\\
Let us check whether the subtree $t|\overline{21}=
g(f_1(x_3,f_1(x_4,x_3)))$ is essential for $(t,\A).$ Its set of
independent subtrees is prefix determined by the positions
$\overline 1$ and $\overline{22}.$ The set of inputs of these
subtrees is $\{x_1,x_2\}.$ All possible assignments for our checking
are sixteen (?). We will calculate the automaton's work only for two
of these assignments $\gamma_1(x_1,x_2,x_3,x_4)= (0,0,0,1)$ and
$\gamma_2(x_1,x_2,x_3,x_4)= (0,0,1,1).$ We obtain the following
equations:
$\A(\gamma_1,t|_{\overline{11}})=\A(\gamma_2,t|_{\overline{11}})=q_0,$
$\A(\gamma_1,t|_{\overline{221}})=\A(\gamma_2,t|_{\overline{221}})=q_0$
and
$\A(\gamma_1,t|_{\overline{1}})=\A(\gamma_2,t|_{\overline{1}})=q_1,$
$\A(\gamma_1,t|_{\overline{22}})=\A(\gamma_2,t|_{\overline{22}})=q_1.$
Analogously we have $\A(\gamma_1,t|_{\overline{2112}})=q_0,$
$\A(\gamma_1,t|_{\overline{211}})=q_0,$
$\A(\gamma_1,t|_{\overline{21}})=q_1,$ and
$\A(\gamma_2,t|_{\overline{2112}})=q_1,$
$\A(\gamma_2,t|_{\overline{211}})=q_1,$
$\A(\gamma_2,t|_{\overline{21}})=q_0.$
\\
In the same way we obtain
 \\
$\A(\gamma_1,t|_{\overline{2}})=q_1,$
$\A(\gamma_1,t|_{\overline{0}})=q_1,$ and
\\
$\A(\gamma_2,t|_{\overline{2}})=q_1,$
$\A(\gamma_2,t|_{\overline{0}})=q_1.$
\\
In the same way it can be checked that for all pairs of assignments
with equal values on $\{x_1,x_2\}$ the automaton will stop in the
same states. Hence $t|_{\overline{21}}$ is not essential subtree for
$(t,\A).$
\\
Now, let us check whether $t|_{\overline{11}}$ is essential.
Consider the following assignments: $\gamma_3(x_1,x_2,x_3,x_4)=
(0,1,1,0)$ and $\gamma_4(x_1,x_2,x_3,x_4)= (1,1,1,0).$ It is easy to
see that $\A(\gamma_3,t|_{\overline{11}})=q_0,$ and
$\A(\gamma_4,t|_{\overline{11}})=q_1.$
\\
So, we obtain $\A(\gamma_3,t)=q_1,$ and $\A(\gamma_4,t)=q_0.$ Thus
$t|_{\overline{11}}$ is essential for $(t,\A).$
\\
Finally it can be checked that for each possible assignment
$\gamma\in Ass(X,\F_0)$ it holds
$\A(\gamma,t)=\A(\gamma,t|_{\overline{1}}).$
\\
This equation implies that the resulting states when $\A$ running
over $t$ are fully determined by the corresponding states when
running over the tree $t|_{\overline 1}=g(f_1(x_1,x_2))$  with the
graph given on Figure \ref{f2}.

\begin{figure}
  \includegraphics[width=10cm]{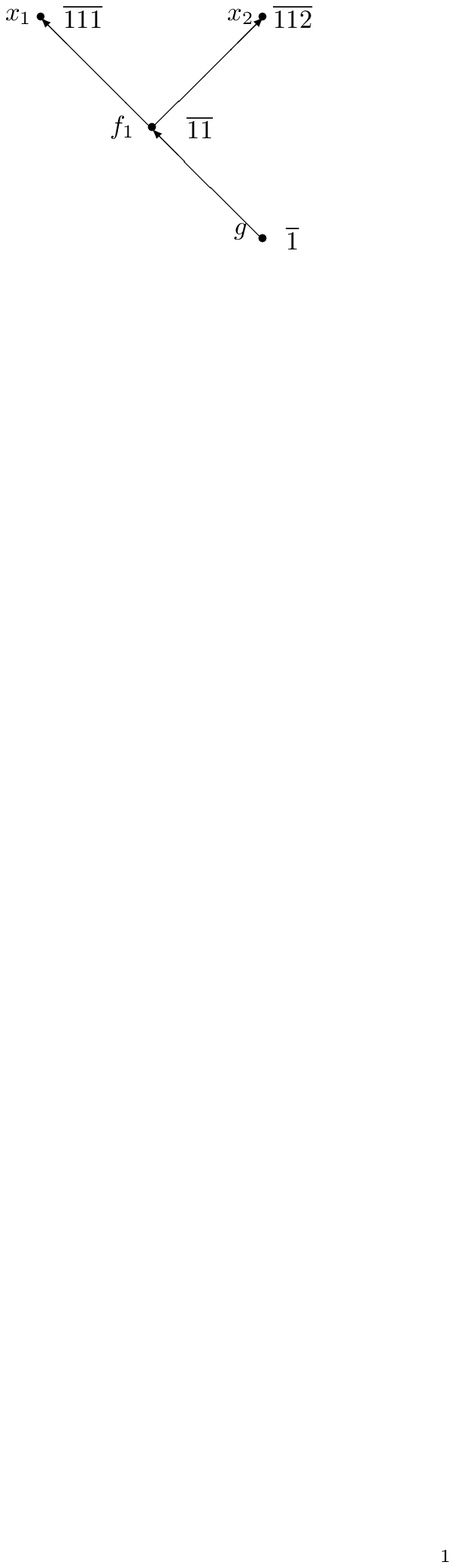}\\
  \caption{~~}\label{f2}
\end{figure}

 It is clear that the runs of $\A$ over $t|_{\overline
1}$ can be calculate more quick than the runs over $t.$
\end{e00}
\begin{t00}\label{t1}\rm
The set $PEss(t,\A)$ is prefix closed.
\end{t00}
\Pr\rm We have to prove that if $t|_{\overline p}\in TEss(t,{ \A})$
then each subtree $t|_{\overline q}$ with $\overline q\prec
\overline p$ is essential for $(t,{\A})$. \ Let
$\gamma_1,\gamma_2\in Ass(X,{\F}_0)$ be two assignments, such that
$\gamma_1(x_j)=\gamma_2(x_j),\quad \mbox{\rm for}\quad x_j\in
X\setminus Var(t|_{\overline p})$ with
$$\A(\gamma_1,t|_{\overline
p})\neq\A(\gamma_2,t|_{\overline p})$$ and
$${\A}(\gamma_1,t)\neq {\A}(\gamma_2,t).$$
Let $\overline q$ be a prefix of $\overline p$ i.e. $\overline
q\prec \overline p.$
\\
At first, suppose  $\A(\gamma_1,t|_{\overline
q})\neq\A(\gamma_2,t|_{\overline q}).$ Then $t|_{\overline q}\in
TEss(t,\A)$ because of $ Var(t|_{\overline p})\subset
Var(t|_{\overline q}).$
\\
 Secondly, let us suppose
\begin{equation}\label{eq111}
\A(\gamma_1,t|_{\overline q})=\A(\gamma_2,t|_{\overline q}).
\end{equation}
From $\A(\gamma_1,t|_{\overline p})\neq\A(\gamma_2,t|_{\overline
p})$ it follows that $Y=Var(t|_{\overline q})\setminus
Var(t|_{\overline p})\neq \emptyset.$ Consider the assignments
$\gamma\in Ass(X\setminus Var(t|_{\overline p}), \F_0) $ and
$\alpha, \beta\in Ass(Var(t|_{\overline p}),\F_0)$ defined as
follows $\gamma(x_i)=\gamma_1(x_i)=\gamma_2(x_i)$ for all $x_i\in
X\setminus Var(t|_{\overline p}), $ $\alpha(x_j)=\gamma_1(x_j)$ and
$\beta(x_j)=\gamma_2(x_j)$ for all $x_j\in Var(t|_{\overline p}).$
\\
Clearly $\gamma\alpha=\gamma_1$ and $\gamma\beta=\gamma_2.$ From
(\ref{eq111}) we have $\A(\alpha,t|_{\overline
q})=\A(\beta,t|_{\overline q}).$ Hence
$$
\A(\gamma_1,t)=\A(\gamma\alpha,t)=\A(\gamma,\A(\alpha,t|_{\overline
q}))=\A(\gamma,\A(\beta,t|_{\overline
q}))=\A(\gamma\beta,t)=\A(\gamma_2,t).
$$
This contradicts to $t|_{\overline p}\in TEss(t,{ \A}).$ \fbx
\begin{c00}\rm \cite{s1}
If $x_i$ is essential input variable for $(t,\A)$ then
there exists a strong
chain
$x_i=t_1\lhd t_2\lhd \ldots \lhd t_k\unlhd t$
such that $x_i\in Ess(t_j,{ \A})$ for $j=1,\ldots,k.$
\end{c00}
\begin{c00}\rm
If $t|_{\overline p}$ is essential subtree for $(t,\A)$ and
$\gamma_1,\gamma_2\in Ass(X,\F_0)$ are as in Theorem \ref{t1} then
$\A(\gamma_1,t|_{\overline q})\neq\A(\gamma_2,t|_{\overline q})$ for
each $\overline q,\ \overline q\prec \overline p.$
\end{c00}
\begin{c00}\rm
The set $P=\{\overline p \in Pos(t)\ |\ {\overline p}\notin
PEss(t,\A)\}$ is prefix determined w.r.t. $Pos(t).$
\end{c00}
\Pr\rm Suppose that $\overline q\in Pos(t),\ \overline
q\prec\overline r,\ \overline r\in Pos(t)$ and $\overline q\notin
PEss(t,\A).$
 If we suppose that
$\overline r\in PEss(t,\A)$ then we obtain $\overline q\in
PEss(t,\A)$ which is a contradiction. \fbx
\\
  Two sets of subtrees $Y$ and $Z$ are called {\it independent}
($Y\rightleftharpoons Z$) if for each $u\in Y$ and for each $v\in
Z$ the inclusions $Y\subset Ind(v)$ and $Z\subset Ind(u)$ are
held.
\begin{d00}\rm\label{d13}
Let $t\in W_{\tau}(X)$ and $\A$ be an automaton. A set $Y\subseteq
TEss(t,\A)$ is called {\it separable} for $t$ and $\A$ w.r.t. a
set $Z\subseteq TEss(t,\A),$ \ $Y\rightleftharpoons Z$ if there is
an assignment $\gamma$ on $\cup_{s\in Z}Var(s)\setminus\cup_{r\in
Y}Var(r)$ such that $Y\subseteq TEss(\gamma(t),\A).$
\end{d00}
 The set of all separable sets of subtrees for $t$ and $\A$ w.r.t. $Z$ will be denoted
by $TSep(t,\A,Z).$ When $Y$ is separable for $t$ and $\A$ w.r.t.
$Z=Ind(Y)$ where $Ind(Y)=\cup_{s\in Y}Ind(s)$ then the set $Y$ is called {\it separable} for $t$ and
$\A$ and the set of such $Y$ will be denoted by
$TSep(t,\A).$
\\
When a set of essential subtrees is not separable, it will be called {\it inseparable}.
\\
It is not difficult  to see that if \ $\forall \gamma\in
Ass(X,{\F}_0)\quad
 {\A}(\gamma,t')={\A}(\gamma,t) $
then
$$
TEss(t,{\A})=TEss(t',{\A}).
$$
\begin{t00}\label{t2}\rm
If $t'$ is an essential subtree of $t$ and  $\forall \gamma\in
Ass(X,{\F}_0)$
\begin{equation}\label{e1}
 {\A}(\gamma,t')={\A}(\gamma,t)
\end{equation}
then each subtree  $s\in Ind(t')$ with $Var(s)\setminus
Var(t')\neq\emptyset,$ is  nonessential  for $(t,\A)$.
\end{t00}
\Pr\rm Suppose there is a subtree $s$ of $t$ with $s\in
Ind(t')\cap TEss(t,\A).$ Then there are two assignments
$\gamma_1,\gamma_2\in Ass(X,\F_0)$ such that $\A(\gamma_1,s)\neq
\A(\gamma_2,s),\quad \mbox{\rm and}\quad
\gamma_1(x_j)=\gamma_2(x_j) \quad\mbox{\rm for all}\quad  x_j\in
X\setminus Var(s)$
 with
\begin{equation}\label{e2}
{\A}(\gamma_1,t)\neq {\A}(\gamma_2,t).
\end{equation}
\\
Now, $s\in Ind(t')$  implies that
\begin{equation}\label{e3}
{\A}(\gamma_1,t')= {\A}(\gamma_2,t').
\end{equation}
On the other side by (\ref{e1}) and (\ref{e3}) we obtain
$${\A}(\gamma_1,t)={\A}(\gamma_1,t')= {\A}(\gamma_2,t')={\A}(\gamma_2,t)$$
which contradicts to (\ref{e2}). \fbx
\begin{t00}\label{t3}\rm
If $Y\in TSep(t,\A)$ then for every subtree $s\in Y$ there exists
at least one strong chain $s=t_1\lhd t_2\lhd \ldots \lhd t_k\unlhd
t$ such that $s\in TEss(t_j,\A)$ for $j=1,\ldots,k.$
\end{t00}
\Pr\rm Let $t=f(s_1,\ldots,s_n).$ By $s\in TEss(t,\A)$ it follows
that there are two assignments $\gamma_1,\gamma_2\in Ass(X,\F_0)$
such that
$$\A(\gamma_1,s)\neq \A(\gamma_2,s),\ \ \mbox{\rm
and}\ \ \gamma_1(x_j)=\gamma_2(x_j)
 \ \ \mbox{\rm for all}\ \  x_j\in  Var(t|_{\overline q}),\ t|_{\overline q}\in
 Ind(s)$$
 with
${\A}(\gamma_1,t)\neq {\A}(\gamma_2,t).$
\\
At first, if $Depth(t)-Depth(s)=1$ then the chain $s\unlhd t$ is strong and the
theorem is proved in this case.
\\
Secondly, let us assume $Depth(t)-Depth(s)\geq 2$.
Suppose that $s\notin TEss(s_j,\A)$
for all $j,\quad j\in\{1,\ldots,n\}.$
This implies that $\A(\gamma_1,s_j)=\A(\gamma_2,s_j),
\quad j=1,\ldots,n.$
Let us calculate again $\A(\gamma_1,t)$   and $\A(\gamma_2,t).$
So, we have
\\
$\A(\gamma_1,t)=
\Delta_n(f,\A(\gamma_1,s_1),
\ldots,\A(\gamma_1,s_n))=
\Delta_n(f,\A(\gamma_2,s_1),
\ldots,\A(\gamma_2,s_n))=\A(\gamma_2,t).$
This is a contradiction. Hence there exists a
subterm $s_j,\ j\in\{1,\ldots,n\}$ of $t$
 such that $s\in TEss(s_j,\A).$
Let us set $t_k=s_j.$ Clearly $t_k\lhd t$ is a strong chain. We
can repeat this procedure for $t_k$ instead of $t$ and we shell
get a subterm $t_{k-1}$ of $t_k$ such that $t_{k-1}\lhd t_k\lhd t$
is a strong chain and $s\in TEss(t_{k-1},\A).$ This process can be
continued until we obtain a term $t_1$ with
$Depth(t_2)-Depth(s)=1$ and $s\in TEss(t_1,\A).$ Obviously if we
set $s=t_1$ then $s=t_1\lhd t_2\lhd \ldots \lhd t_k\lhd t$ is a
strong chain with  $s\in TEss(t_j,\A),\quad j=1,\ldots,k.$ \fbx

\end{document}